\documentclass[11pt,a4paper]{article}
\usepackage[utf8]{inputenc}

\usepackage[T1]{fontenc}
\usepackage{fullpage}
\usepackage{amsmath}
\usepackage[dvipsnames]{xcolor}
\usepackage{subcaption,tikz}
\usetikzlibrary{matrix,arrows,decorations.pathmorphing}
\usepackage{amssymb,mathrsfs}
\usepackage{bbm}
\usepackage{xr}
\usepackage{comment}
\usepackage{amsthm}
\usepackage[all,cmtip]{xy}
\usepackage[toc,page]{appendix}

\usepackage{subcaption}
\usepackage{tabularx}
\newcolumntype{Y}{>{\centering\arraybackslash}X}
\usepackage{url}
\usepackage[british]{babel}
\usepackage{courier}
\usepackage{enumerate}
\usepackage[T1]{fontenc}
\usepackage{listings}
\usepackage{algorithm}
\usepackage[noend]{algpseudocode}
\usepackage{authblk}
\usepackage{cite}
\usepackage[normalem]{ulem}
\usepackage{siunitx} 
\usepackage[pdftex,breaklinks,linkbordercolor={0.8 0.8 1.0},
citebordercolor={0.8 1.0 0.8},urlbordercolor={0.8 1 1}]{hyperref}

\title{\textit{Structural heterogeneity}: a topological characteristic to track the time evolution of soft matter systems}
\date{}

\author[1]{Ingrid~Membrillo~Solis\footnote{Corresponding author}}
\author[2]{Tetiana~Orlova}
\author[3]{Karolina~Bednarska}
\author[3]{Piotr~Lesiak}
\author[3]{Tomasz~R.~Woli\'{n}ski}
\author[1]{Giampaolo~D'Alessandro}
\author[1]{Jacek~Brodzki}
\author[2]{Malgosia~Kaczmarek}
\affil[1]{Mathematical Sciences, University of Southampton,  Southampton SO17~1BJ, UK}
\affil[2]{Physics and Astronomy, University of Southampton,  Southampton SO17~1BJ, UK}
\affil[3]{Faculty of Physics, Warsaw University of Technology, Koszykowa 75, 00-662 Warszawa, Poland}

\begin{document}
\maketitle

\begin{abstract}
We introduce \textit{structural heterogeneity}, a new topological characteristic for semi-ordered materials that captures their degree of organisation at a mesoscopic level and tracks their time-evolution, ultimately detecting the order-disorder transition at the microscopic scale. 
Such quantitative characterisation of a complex, soft matter system has not yet been achieved with any other method.
We show that structural heterogeneity can track structural changes in a liquid crystal nanocomposite, reveal the effect of confined geometry on the nematic-isotropic and isotropic-nematic phase transitions, and uncover physical differences between these two processes. The system used in this work is representative of a class of composite nanomaterials, partially ordered and with complex structural and physical behaviour,  where their precise characterisation poses significant challenges. Our newly developed analytic framework can provide both a qualitative and a quantitative characterisations of the dynamical behaviour of a wide range of semi-ordered soft matter systems.
\end{abstract}

 Many soft composite and biological materials present very complex phase dynamics \cite{Jabbari2013, Hwang2016, Patra2014, Panter2019, Li2016, Jacobs2017, Chu2020, Wales2018, Charbonneau2018} that require powerful analytic methods for their accurate and quantitative characterisation.  Persistent homology, a tool from topological data analysis \cite{ELZ02,E08,ZC05}, has recently emerged as an effective, quantitative method to reveal structural and morphological features in various soft materials, such as granular samples \cite{ardanza2014,saadatfar2017}, silica glasses \cite{hiraoka2016}, glassy polymers \cite{ichinomiya2017}, living cells, tissues~\cite{lawson2019, oyama2019, teramoto2020, ferri2017, qaiser2016, Nicponski2020} and  biological objects \cite{Amezquita2020, McGuirl2020}. Although this topological method has been successfully used to characterise these kinds of materials, to the best of our knowledge, neither persistent homology nor any other analytic method have been used to characterise the dynamical behaviour of structured soft matter systems using optical experimental data. 

In this paper we introduce \emph{structural heterogeneity}, a persistent homology-based characteristic for semi-organised soft matter systems. Structural heterogeneity allows one to measure the deviation of a soft matter system from being in a homogeneous or uniform state at a mesoscopic scale. In this work, we use structural heterogeneity to analyse the time-evolution of a nematic liquid crystal doped with gold nanoparticles~\cite{Lesiak2019} in phase transition processes. This system has an intrinsically complex dynamics that makes it a non-trivial example of an evolving soft-matter system: the plasmonic nanoparticles are embedded in a non-isotropic complex fluid that organises them in a periodic manner as it evolves from the nematic to the isotropic phase, and retains them in this configuration when the process is reversed. In addition, this composite fluid is constrained by the cylindrical geometry of the capillary that contains it. At the same time, the system is quasi one-dimensional, which simplifies  data acquisition and data analysis. Furthermore, we obtain a representation of each topological descriptor, associated to a particular state of the system, as a point in a Euclidean space. This representation allows us to characterise algorithmically the dynamical behaviour of the system. We show that our topological methods detect distinct temperature-induced macroscopic states, with different degrees of order.

An important outcome of our analysis is the development of a persistent homology-based framework to quantify the structural changes experienced by a soft matter system during a thermodynamical process. The novelty of our framework is that persistent homology is used to reveal the structural organisation at the mesoscopic scale, when the analysed system consists of a number of micron-size molecular ensembles with different order parameter~\cite{Lesiak2019}. Even though persistent homology has been used in previous works to characterise the order phase transitions in spin lattices~\cite{tran2021, donato2016, cole2020}, in none of them persistent homology was used to analyse  the dynamical behaviour of the materials at the supramolecular level.

In the context of liquid crystals, our results can potentially lead to more elaborate free energy landscapes for nanoparticle-loaded materials and to provide a new perspective on the Landau-de Gennes theory of phase transitions between nematic and isotropic phases. The  importance of our pipeline is, however, far more general: it shows that persistent homology can be used as a powerful quantitative method to characterise the complex phase dynamics of non-homogeneous soft materials.

\begin{figure}
    \centering

\begin{subfigure}[b]{0.25\textwidth}
\centering
\includegraphics[width=0.9\textwidth]{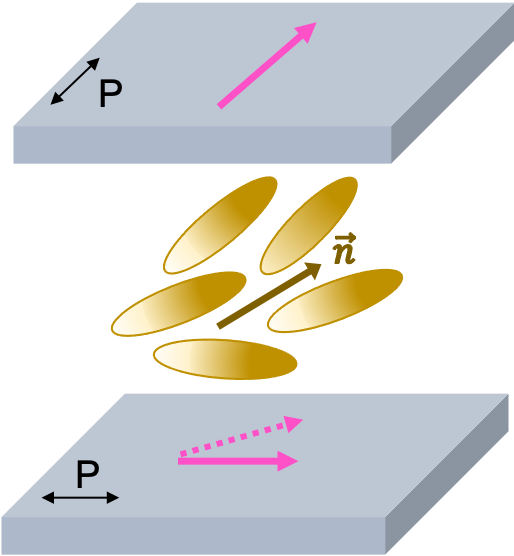}
\caption{}
\end{subfigure}\hfill%
\begin{subfigure}[b]{0.4\textwidth}
\centering
\begin{tikzpicture}[scale=0.68]
\fill [black] (0,1) rectangle (1,4) (1,3) rectangle (2,4) (2,1) rectangle (3,3) (4,1) rectangle (5,3) (5,0) rectangle (6,1);
\fill [black!83] (0,0) rectangle (2,1) (1,1) rectangle (2,2) (2,3) rectangle (4,4) (4,0) rectangle (5,1) (6,0) rectangle (8,1);
\fill [black!67] (2,0) rectangle (4,1) (4,3) rectangle (5,4) (7,1) rectangle (8,4);
\fill [black!50] (3,1) rectangle (4,3) (5,3) rectangle (7,4);
\fill [black!33] (6,1) rectangle (7,3);
\fill [black!17] (5,1) rectangle (6,2);
\draw [very thick,blue] (0,0) rectangle (8,4);
\draw [xshift=0.5cm,yshift=3.5cm,x=0.1cm,ultra thick,magenta] (-0.3,0) -- (0.3,0);
\draw [xshift=0.5cm,yshift=2.5cm,x=0.1cm,ultra thick,magenta] (-0.3,0) -- (0.3,0);
\draw [xshift=0.5cm,yshift=1.5cm,x=0.1cm,ultra thick,magenta] (-0.3,0) -- (0.3,0);
\draw [xshift=0.5cm,yshift=0.5cm,x=0.25cm,ultra thick,magenta] (-0.3,0) -- (0.3,0);
\draw [xshift=1.5cm,yshift=3.5cm,x=0.1cm,ultra thick,magenta] (-0.3,0) -- (0.3,0);
\draw [xshift=1.5cm,yshift=2.5cm,x=1.0cm,ultra thick,magenta] (-0.3,0) -- (0.3,0);
\draw [xshift=1.5cm,yshift=1.5cm,x=0.25cm,ultra thick,magenta] (-0.3,0) -- (0.3,0);
\draw [xshift=1.5cm,yshift=0.5cm,x=0.25cm,ultra thick,magenta] (-0.3,0) -- (0.3,0);
\draw [xshift=2.5cm,yshift=3.5cm,x=0.25cm,ultra thick,magenta] (-0.3,0) -- (0.3,0);
\draw [xshift=2.5cm,yshift=2.5cm,x=0.1cm,ultra thick,magenta] (-0.3,0) -- (0.3,0);
\draw [xshift=2.5cm,yshift=1.5cm,x=0.1cm,ultra thick,magenta] (-0.3,0) -- (0.3,0);
\draw [xshift=2.5cm,yshift=0.5cm,x=0.4cm,ultra thick,magenta] (-0.3,0) -- (0.3,0);
\draw [xshift=3.5cm,yshift=3.5cm,x=0.25cm,ultra thick,magenta] (-0.3,0) -- (0.3,0);
\draw [xshift=3.5cm,yshift=2.5cm,x=0.55cm,ultra thick,magenta] (-0.3,0) -- (0.3,0);
\draw [xshift=3.5cm,yshift=1.5cm,x=0.55cm,ultra thick,magenta] (-0.3,0) -- (0.3,0);
\draw [xshift=3.5cm,yshift=0.5cm,x=0.4cm,ultra thick,magenta] (-0.3,0) -- (0.3,0);
\draw [xshift=4.5cm,yshift=3.5cm,x=0.4cm,ultra thick,magenta] (-0.3,0) -- (0.3,0);
\draw [xshift=4.5cm,yshift=2.5cm,x=0.1cm,ultra thick,magenta] (-0.3,0) -- (0.3,0);
\draw [xshift=4.5cm,yshift=1.5cm,x=0.1cm,ultra thick,magenta] (-0.3,0) -- (0.3,0);
\draw [xshift=4.5cm,yshift=0.5cm,x=0.25cm,ultra thick,magenta] (-0.3,0) -- (0.3,0);
\draw [xshift=5.5cm,yshift=3.5cm,x=0.55cm,ultra thick,magenta] (-0.3,0) -- (0.3,0);
\draw [xshift=5.5cm,yshift=2.5cm,x=1.0cm,ultra thick,magenta] (-0.3,0) -- (0.3,0);
\draw [xshift=5.5cm,yshift=1.5cm,x=0.85cm,ultra thick,magenta] (-0.3,0) -- (0.3,0);
\draw [xshift=5.5cm,yshift=0.5cm,x=0.1cm,ultra thick,magenta] (-0.3,0) -- (0.3,0);
\draw [xshift=6.5cm,yshift=3.5cm,x=0.55cm,ultra thick,magenta] (-0.3,0) -- (0.3,0);
\draw [xshift=6.5cm,yshift=2.5cm,x=0.7cm,ultra thick,magenta] (-0.3,0) -- (0.3,0);
\draw [xshift=6.5cm,yshift=1.5cm,x=0.7cm,ultra thick,magenta] (-0.3,0) -- (0.3,0);
\draw [xshift=6.5cm,yshift=0.5cm,x=0.25cm,ultra thick,magenta] (-0.3,0) -- (0.3,0);
\draw [xshift=7.5cm,yshift=3.5cm,x=0.4cm,ultra thick,magenta] (-0.3,0) -- (0.3,0);
\draw [xshift=7.5cm,yshift=2.5cm,x=0.4cm,ultra thick,magenta] (-0.3,0) -- (0.3,0);
\draw [xshift=7.5cm,yshift=1.5cm,x=0.4cm,ultra thick,magenta] (-0.3,0) -- (0.3,0);
\draw [xshift=7.5cm,yshift=0.5cm,x=0.25cm,ultra thick,magenta] (-0.3,0) -- (0.3,0);
\node at (4,4.5) {$\mathcal{X}$};
\end{tikzpicture}

\scriptsize
\medskip
\begin{tikzpicture}
\foreach \x in {0,...,100} \fill [white!\x!black] (\x*0.054,1) rectangle (\x*0.054+0.07,1.5);
\node [below] at (0,1) {0};
\node [below] at (0.9,1) {43};
\node [below] at (1.8,1) {85};
\node [below] at (2.7,1) {128};
\node [below] at (3.6,1) {170};
\node [below] at (4.5,1) {213};
\node [below] at (5.4,1) {255};
\end{tikzpicture}
\caption{}
\end{subfigure}\hfill%
\begin{subfigure}[b]{0.32\textwidth}
\stepcounter{subfigure}
\centering
\begin{tikzpicture}[scale=0.6]
\draw [thick,->] (-1,0) -- (6,0) node [below] {\scriptsize birth};
\draw [thick,->] (0,-1) -- (0,7) node [left] {\scriptsize death};
\draw [thick,loosely dotted] (0,6) node [left] {\scriptsize 255} -- (3,6) (0,3) node [left] {\scriptsize 128} -- (2,3) (1,1) -- (1,0) node [below] {\scriptsize 43} (2,2) -- (2,0) node [below] {\scriptsize 85} (3,3) -- (3,0) node [below] {\scriptsize 128};
\fill [violet!70] (1,6) circle (0.15) node [above] {\scriptsize\:\:{}$(43,255)$};
\draw [very thick,violet!70,densely dotted] (1,6) -- (1,1);
\fill [green!80!black] (3,6) circle (0.15) node [above] {\scriptsize\qquad\quad$(128,255)$};
\draw [very thick,green!80!black,densely dotted] (3,6) -- (3,3);
\fill [cyan] (2,3) circle (0.15) node [above] {\scriptsize $(85,128)$};
\draw [very thick,cyan,densely dotted] (2,3) -- (2,2);
\draw [very thick] (0,0) node [below left] {\scriptsize 0} -- (6,6);
\end{tikzpicture}
\caption{}
\end{subfigure}

\bigskip

\begin{subfigure}[b]{\textwidth}
\addtocounter{subfigure}{-2}
\centering
\begin{tikzpicture}[scale=0.25,rotate=90]
\begin{scope}[yshift=0.5cm,rotate=-90]
\fill [black] (0,1) rectangle (1,4) (1,3) rectangle (2,4) (2,1) rectangle (3,3) (4,1) rectangle (5,3) (5,0) rectangle (6,1);
\fill [black!83] (0,0) rectangle (2,1) (1,1) rectangle (2,2) (2,3) rectangle (4,4) (4,0) rectangle (5,1) (6,0) rectangle (8,1);
\fill [black!67] (2,0) rectangle (4,1) (4,3) rectangle (5,4) (7,1) rectangle (8,4);
\fill [black!50] (3,1) rectangle (4,3) (5,3) rectangle (7,4);
\fill [black!33] (6,1) rectangle (7,3);
\fill [black!17] (5,1) rectangle (6,2);
\fill [white] (1,2) rectangle (2,3) (5,2) rectangle (6,3);
\node at (4,5) {$\mathcal{X}(255)$};
\end{scope}
\begin{scope}[yshift=9.5cm,rotate=-90]
\fill [black] (0,1) rectangle (1,4) (1,3) rectangle (2,4) (2,1) rectangle (3,3) (4,1) rectangle (5,3) (5,0) rectangle (6,1);
\fill [black!83] (0,0) rectangle (2,1) (1,1) rectangle (2,2) (2,3) rectangle (4,4) (4,0) rectangle (5,1) (6,0) rectangle (8,1);
\fill [black!67] (2,0) rectangle (4,1) (4,3) rectangle (5,4) (7,1) rectangle (8,4);
\fill [black!50] (3,1) rectangle (4,3) (5,3) rectangle (7,4);
\fill [black!33] (6,1) rectangle (7,3);
\fill [black!17] (5,1) rectangle (6,2);
\draw [violet!70,ultra thick] (1,2) rectangle (2,3);
\draw [green!80!black,ultra thick] (5,2) rectangle (6,3);
\node at (4,5) {$\mathcal{X}(213)$};
\end{scope}
\begin{scope}[yshift=18.5cm,rotate=-90]
\fill [black] (0,1) rectangle (1,4) (1,3) rectangle (2,4) (2,1) rectangle (3,3) (4,1) rectangle (5,3) (5,0) rectangle (6,1);
\fill [black!83] (0,0) rectangle (2,1) (1,1) rectangle (2,2) (2,3) rectangle (4,4) (4,0) rectangle (5,1) (6,0) rectangle (8,1);
\fill [black!67] (2,0) rectangle (4,1) (4,3) rectangle (5,4) (7,1) rectangle (8,4);
\fill [black!50] (3,1) rectangle (4,3) (5,3) rectangle (7,4);
\fill [black!33] (6,1) rectangle (7,3);
\draw [violet!70,ultra thick] (1,2) rectangle (2,3);
\draw [green!80!black,ultra thick] (5,1) rectangle (6,3);
\node at (4,5) {$\mathcal{X}(170)$};
\end{scope}
\begin{scope}[yshift=27.5cm,rotate=-90]
\fill [black] (0,1) rectangle (1,4) (1,3) rectangle (2,4) (2,1) rectangle (3,3) (4,1) rectangle (5,3) (5,0) rectangle (6,1);
\fill [black!83] (0,0) rectangle (2,1) (1,1) rectangle (2,2) (2,3) rectangle (4,4) (4,0) rectangle (5,1) (6,0) rectangle (8,1);
\fill [black!67] (2,0) rectangle (4,1) (4,3) rectangle (5,4) (7,1) rectangle (8,4);
\fill [black!50] (3,1) rectangle (4,3) (5,3) rectangle (7,4);
\draw [violet!70,ultra thick] (1,2) rectangle (2,3);
\draw [green!80!black,ultra thick] (5,1) rectangle (7,3);
\node at (4,5) {$\mathcal{X}(128)$};
\end{scope}
\begin{scope}[yshift=36.5cm,rotate=-90]
\fill [black] (0,1) rectangle (1,4) (1,3) rectangle (2,4) (2,1) rectangle (3,3) (4,1) rectangle (5,3) (5,0) rectangle (6,1);
\fill [black!83] (0,0) rectangle (2,1) (1,1) rectangle (2,2) (2,3) rectangle (4,4) (4,0) rectangle (5,1) (6,0) rectangle (8,1);
\fill [black!67] (2,0) rectangle (4,1) (4,3) rectangle (5,4) (7,1) rectangle (8,4);
\draw [violet!70,ultra thick] (1,2) rectangle (2,3);
\draw [cyan,ultra thick] (3,1) rectangle (4,3);
\node at (4,5) {$\mathcal{X}(85)$};
\end{scope}
\begin{scope}[yshift=45.5cm,rotate=-90]
\fill [black] (0,1) rectangle (1,4) (1,3) rectangle (2,4) (2,1) rectangle (3,3) (4,1) rectangle (5,3) (5,0) rectangle (6,1);
\fill [black!83] (0,0) rectangle (2,1) (1,1) rectangle (2,2) (2,3) rectangle (4,4) (4,0) rectangle (5,1) (6,0) rectangle (8,1);
\draw [violet!70,ultra thick] (1,2) rectangle (2,3);
\node at (4,5) {$\mathcal{X}(43)$};
\end{scope}
\begin{scope}[yshift=54.5cm,rotate=-90]
\fill [black] (0,1) rectangle (1,4) (1,3) rectangle (2,4) (2,1) rectangle (3,3) (4,1) rectangle (5,3) (5,0) rectangle (6,1);
\node at (4,5) {$\mathcal{X}(0)$};
\end{scope}
\end{tikzpicture}
\caption{}
\end{subfigure}

\begin{subfigure}[b]{0.3\textwidth}
\stepcounter{subfigure}
\centering
\includegraphics[width=\textwidth]{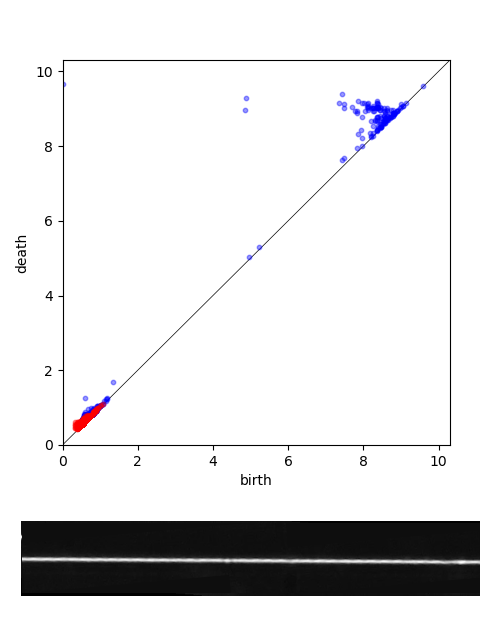}
\caption{}
\end{subfigure}\hfill%
\begin{subfigure}[b]{0.3\textwidth}
\centering
\includegraphics[width=\textwidth]{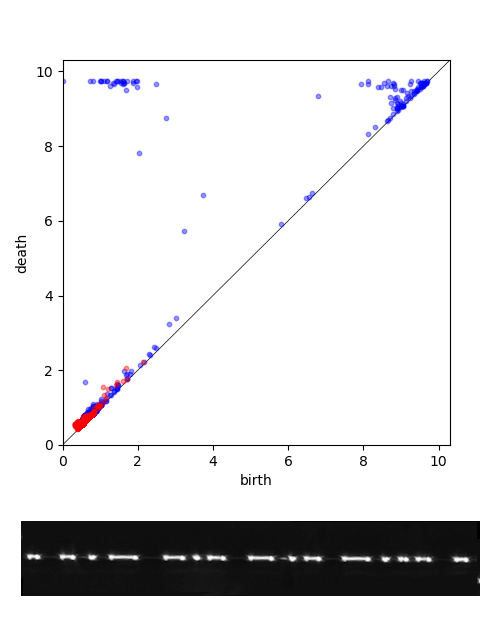}
\caption{}
\end{subfigure}\hfill%
\begin{subfigure}[b]{0.3\textwidth}
\centering
\includegraphics[width=\textwidth]{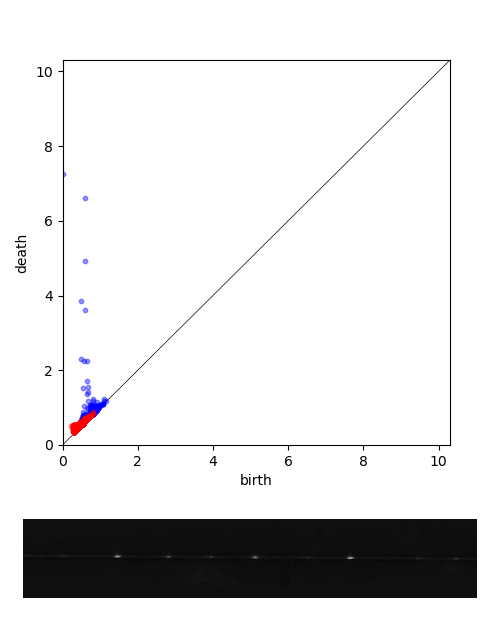}
\caption{}
\end{subfigure}

\caption{\textbf{Construction of topological descriptors.} \textbf{a,} Semi-disordered rod-shaped molecules inside a capillary placed between a pair of $90^{\circ}$-crossed polarisers at $45^{\circ}$ to their axes. The bottom solid magenta arrow represents the final polarisation and amplitude of incident light having passed through the polarisers and the LC; the top solid magenta arrow represents the input polarisation and the intermediate stage is marked in the dashed magenta arrow. \textbf{b,} A grey-scale picture $\mathcal X$ (top) representing a state where two mesophases coexist and the  colour map bar (bottom) given by light intensity of the pixel; the orientation of the magenta bars is determined by the axis of the lower polariser, and the bar lengths correspond to the average intensity of transmitted light. \textbf{c}, Some partial pictures $\mathcal X(i)$ of $\mathcal X$; the 1-loops of the filtration are marked (purple, blue, green).  \textbf{d,} The persistence diagram  of $\mathcal X$, showing three 1-cycles with the colour corresponding to their  loops of pixels associated to them (purple, blue and green ); \textbf{e-g,} Experimental grey-scale picture of the LC filled capillary at different stages in the evolution of the system (bottom) with their corresponding persistence diagrams (top); 0-cycles are marked red in the persistence diagram, whereas 1-cycles are marked blue. }
    \label{fig:filtration}
\end{figure}
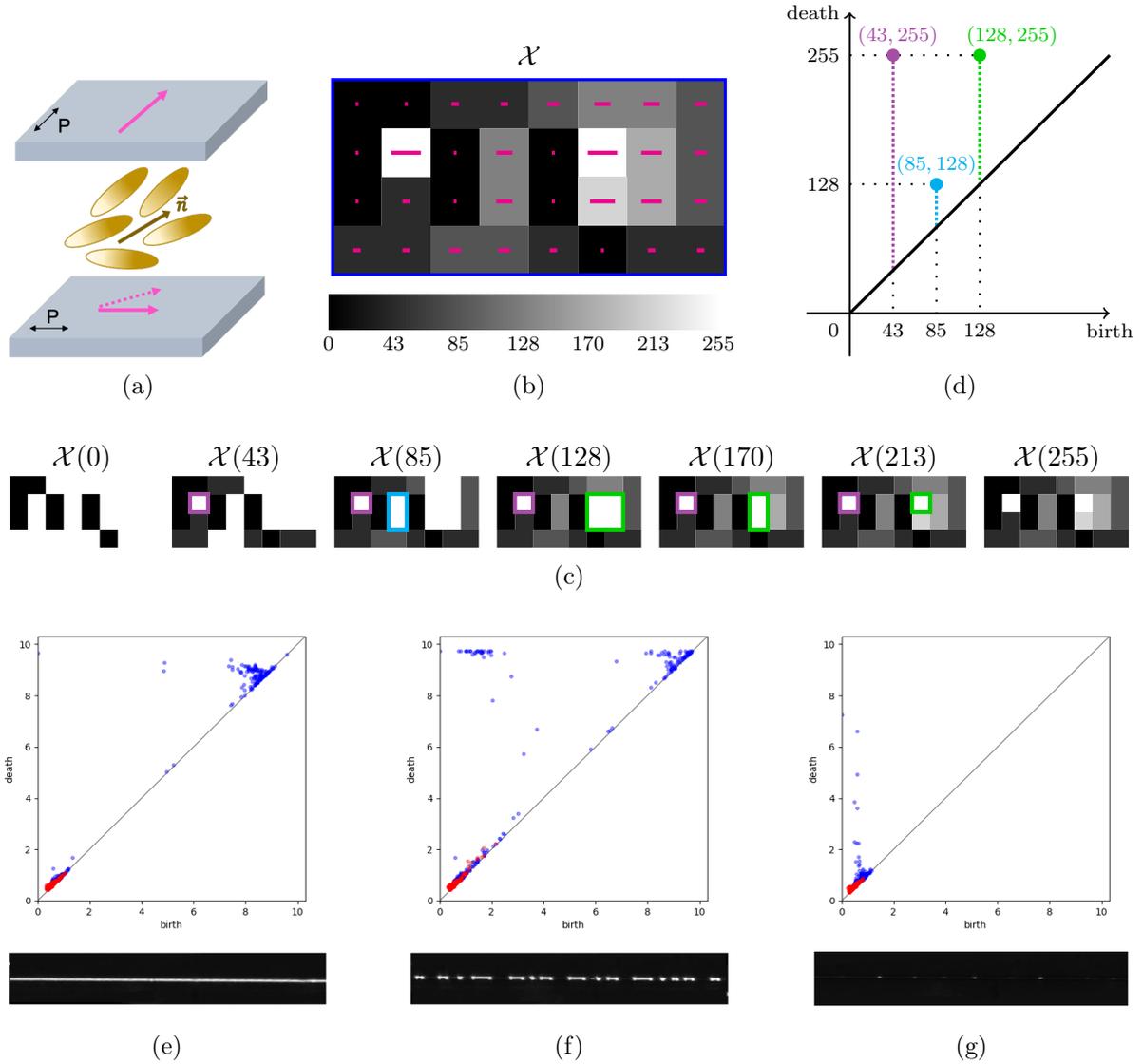

\section*{Structural heterogeneity }

The goal of this section is to define  \emph{structural heterogeneity} (SH) as a new topological characteristic for soft matter systems. SH is based on persistent homology, a topological tool used to extract meaningful information from the shape of the data. To help with the discussion, we start by describing the physical properties and the topological features observed in a toy model of a soft matter system. Fig.~\ref{fig:filtration} a shows a cross section of a confined LC system placed between two cross-polarisers; a grey-scale picture of the whole system is showed in Fig.~\ref{fig:filtration} b, where each pixel (square) represents a 2D projection of Fig.~\ref{fig:filtration} a. In this set-up, light passes through the polariser at the top, interacts with an ensemble of LC molecules, to finally cross the polariser at the bottom. The higher the order of the molecular ensemble is, the higher the intensity value of the pixel. A magenta arrow and a bar represent the polarisation and the intensity of the light. In the picture, there are pixels of high intensity values surrounded by loops of pixels of lower intensity, which are generated by highly ordered molecular ensembles surrounded by loops of highly disordered molecules. This picture might indicate, for instance, that the system is in a state where two distinct mesophases coexist. From this analysis, one can see that the quantification of the topological features present in the images of a semi-organised system might be used to characterise it.

Before defining SH, we give a brief introduction to persistent homology (PH). A more detailed exposition of PH is found in the Supplementary Material (SM). Starting with a grey-scale picture $\mathcal{X}$ as in Fig.~\ref{fig:filtration} b, and a number $i$, with $0 \leq i \leq 255$, we define a partial picture $\mathcal{X}(i)$ as the union of pixels in $\mathcal{X}$ with light intensity not greater than $i$; changing the value of $i$ creates a set of partial pictures parametrised by light intensity. The topological features of $\mathcal X$, are analysed by  keeping track of the appearance and disappearance of the topological features in the partial pictures $\mathcal X(i)$ as $i$ increases, Fig.~\ref{fig:filtration} c. There are two types of topological features to analyse:  connected pieces, or 0-cycles, and loops of pixels, or 1-cycles. Each topological feature $\alpha$ observed in the partial pictures is represented by a point $(b_\alpha, d_\alpha)$ in the Euclidean plane, where $b_\alpha$ is the value $i$ when $\alpha$ appears for the first time, and $d_\alpha$ when it disappears. The values $b_\alpha$ and $d_\alpha$ are called the \emph{birth} and the \emph{death} of $\alpha$, respectively. The collection of all the points $(b_\alpha,d_\alpha)$ representing topological features observed in $\{\mathcal X(i)\}$ is called the \emph{persistence diagram} (PD) of $\mathcal{X}$ (Fig.~\ref{fig:filtration} d). The most relevant physical information is contained in the 1-cycles. The value $d_\alpha-b_\alpha$ is called the \emph{persistence} of the topological feature $\alpha$. The persistence of a 1-cycle measures the difference in order between the inside and the rim of the loop of pixels.  

If $\mathcal{X}$ is a picture representing a physical state of a system, and $\operatorname{PD}(\mathcal{X})$ is its corresponding PD, we define the \emph{structural heterogeneity} (SH) of $\mathcal{X}$ as the sum of the persistence values over all $1$-cycles $\alpha$ in $\operatorname{PD}(\mathcal{X})$,
\[
\operatorname{SH} = \sum_{\text{1-cycles}} |d_\alpha - b_\alpha|. 
\]

SH measures the deviation of a soft matter system from being in a homogeneous or uniform state. We used SH to quantify the deviation of a soft matter system from being in either a totally nematic or a totally isotropic phase during a phase transition: the bigger the number of ordered molecular clusters surrounded by disordered molecular loops, the higher the value of SH.

The analysed systems consisted of  capillaries of various inner diameters, filled with a nematic LC containing uniformly dispersed gold nanospheres \cite{Lesiak2019} (see SM for further details on the experimental set-up). The capillaries were rapidly heated, letting the composite material reach the isotropic state, and then cooled back to the nematic phase. During the heating process, isotropic domains formed and grew in size, expelling nanoparticles into the adjacent nematic zones. This led to the self-assembly of nanospheres into a one-dimensional periodic structure along the capillary. Then the isotropic domains merged, forming an isotropic state with non-homogeneous particle density. During the isotropic-nematic process, nematic domains first nucleated in regions with high nanoparticle concentration, and then expanded to the rest of the capillary. All phase transitions were video-captured under imaging conditions where the nematic and isotropic domains appear bright and dark, respectively. Each video frame was digitised using integer grey-scale levels that implicitly represent the LC order, from 0 (black, isotropic phase) to 255 (white, nematic phase). We obtained a persistence diagram for each video frame of the analysed systems (see SM). In Fig.~\ref{fig:filtration} e-g we show three PDs (top)  along with their corresponding video frames (bottom) at different steps of the nematic-isotropic phase transition.

\begin{figure}
  \centering
  \includegraphics[width=\textwidth]{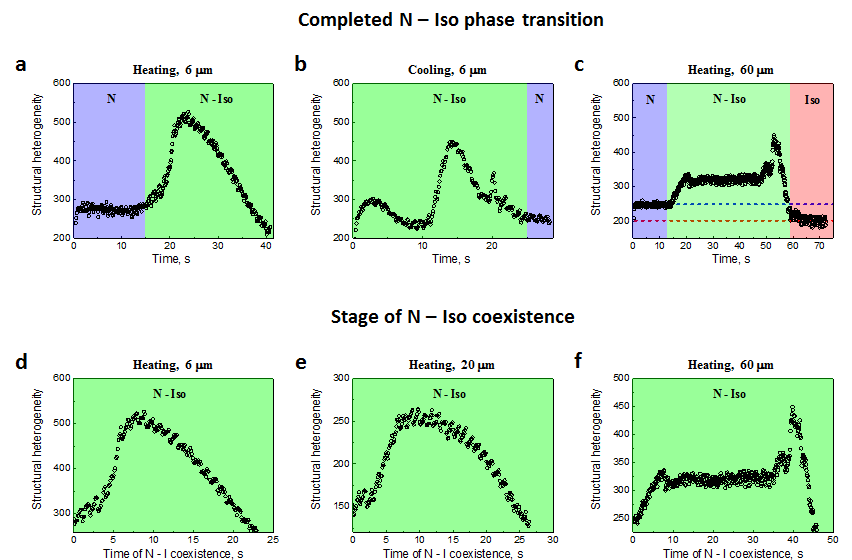}
  \caption{\textbf{Time dynamics of the structural heterogeneity.} In all plots the horizontal axis is the time of the video frames of the capillary and the vertical axis is the corresponding structural heterogeneity. \textbf{a,b,} The structural heterogeneity of the heating and cooling process respectively in a \SI{6}{\micro m} capillary. \textbf{c,} The entire nematic-isotropic transition in a \SI{60}{\micro m} capillary that shows the difference of the structural heterogeneity in the nematic (left) and isotropic (right) state. The nematic, nematic-isotropic  and isotropic states of the LC are highlighted in blue, green and red, respectively. \textbf{d, e, f,} Increase of the time range of coexistence of nematic and isotropic domains in capillary of 6 \si{\micro m}, 20 \si{\micro m} and 60 \si{\micro m}, respectively. }
  \label{fig:norm}
\end{figure}

We start the discussion of the SH of the LC nanocomposites by analysing the SH plot of the nematic-isotropic phase transition in a capillary of 6$\mu$m, Fig.~\ref{fig:norm} a. In the first seconds of the recording, that is, before the phase transition starts, the molecular organisation in the LC is uniform; during this period the SH  looks roughly constant. The formation of isotropic regions, and, hence, the transition of the system from  nematic to a nematic-isotropic phase (N-Iso), corresponds to an increase in SH up to a maximum; a short period of stabilisation in SH follows,  where no new isotropic domains appear and the existing ones increase in size. The width of the N-Iso region (and of the SH plateau) increases with the diameter of the capillary, Fig.~\ref{fig:norm}~a,c. As the phase transition approaches its end, the nematic domains start collapsing until they vanish. The system then becomes homogeneous, reaching an isotropic state. We note that the SH values of the nematic states and those of the isotropic states are different, even though in both cases, the system  looks homogeneous, Fig.~\ref{fig:norm}~c. 

The observed dependence of the phase transition time on the capillary diameter  (Fig.~\ref{fig:norm}~d-f) can be explained by a phenomenon related to  phase separation: the growing area of the isotropic phase is the source of nanoparticle movement at the interface between the nematic and the isotropic phases, locally increasing their concentration in the nematic phase. This movement slows down the rate of the phase transition considerably. Therefore, in Fig.~\ref{fig:norm} c we can see a clear flat area between the beginning and the end of the phase transition. For smaller diameters, this process is shorter because the nanoparticles are moved over shorter distances (the period of the periodic structure depends on the diameter of the capillary \cite{Lesiak2019}). 

The differences between the nematic-isotropic and isotropic-nematic processes are evidenced by the comparison between Fig.~\ref{fig:norm} a and Fig.~\ref{fig:norm} b: in the nematic-isotropic transition we can observe one local maximum whereas in the inverse process we can observe two well-defined maxima. This indicates that SH detects a hysteresis behaviour, in agreement with  experimental observations reported in \cite{Lesiak2019}. To investigate in more detail the physical meaning of the observed maxima in the SH plots, and the differences between the nematic-isotropic and isotropic-nematic processes, we introduce the concept of topological pathway.

\section*{Topological pathways of dynamical processes}
\label{sec:cms}

To start, we use the \emph{bottleneck distance} to quantify the difference between PDs. This is defined as follows. For PDs $D_1$ and $D_2$, which are subsets of the plane $\mathbb{R}^2$, we consider a bijective matching $\phi$ of points of $D_1$ to points of $D_2$. We also allow matching any of the points of $D_1$ or $D_2$ to its nearest point on the $x=y$ line; a matching like that always exists, even if $D_1$ and $D_2$ have different numbers of points. The \emph{cost}, $\text{Cost}(\phi)$,  of a matching $\phi$ of diagrams $D_1$ and 
$D_2$ is the largest distance by which a point in $D_1$ has to be moved to be matched with a point in $D_2$. Then the bottleneck distance is 
\[
d(D_1, D_2) = \text{Smallest $\text{Cost}(\phi)$ taken over the set of all possible matchings $\phi$.
}
\]
If two PDs are close in this distance, so are the topological features of the corresponding video frames \cite{cohen2007}. Since the topological features of a video frame depend on the physical state of the system, the smaller the bottleneck distance between two PDs is, the more similar the states of the system associated to them are.

\begin{figure}
    \centering
    \includegraphics[scale=0.61]{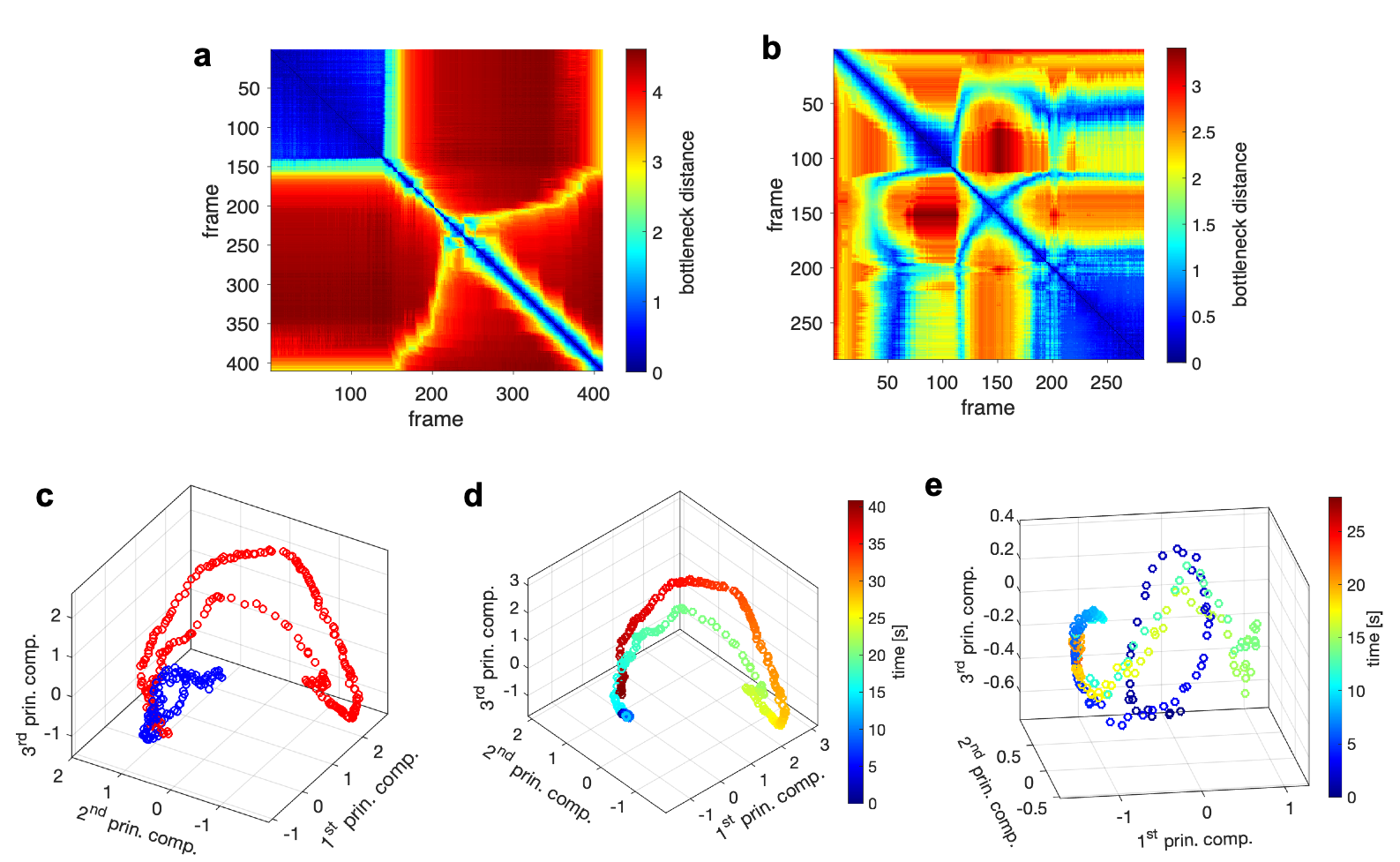}
    \caption{\textbf{Bottleneck distance matrices and 3D representation of the topological pathways of the \SI{6}{\micro m} capillary.} The bottleneck distance matrices associated to \textbf{a,} the nematic-isotropic  and \textbf{b,} the isotropic-nematic phase transitions; here the frame rate is 10 fps. \textbf{c,} Heating (red) and cooling (blue) topological pathways. \textbf{d,} Nematic-isotropic and \textbf{e,} isotropic-nematic topological pathways. In (\textbf{d}) and (\textbf{e}) colour codes time from beginning (blue) to the end (red).}
    \label{fig:dm}
\end{figure}

For each pair of PDs we computed their bottleneck distance.  Fig.~\ref{fig:dm} a,b shows the distance matrices obtained. Here, blue corresponds to pairs of PDs that are practically identical, and red the most different. The distance matrix of the nematic-isotropic transition (Fig.~\ref{fig:dm} a) shows that in  the first 15 s of the recording, approximately, the states of the system are almost identical. Distances between PDs of video frames at $t>15 s$ and at  $t<15 s$ are large. The yellow-cyan area in the center of (Fig.~\ref{fig:dm} a) corresponds to a period in which the distances between PDs start decreasing. From the analysis of the video frames and the SH values, we conclude that, in this short time period, nematic and isotropic phases coexist, and the number of isotropic remains the same. Moreover, the observed period increases with the capillary diameter (see SM). The distance matrices of the nematic-isotropic and the isotropic-nematic phase transitions look substantially different, confirming the differences highlighted by SH.

The bottleneck distance allows us to regard the set of PDs of each dynamical process as a (non-Euclidean) metric space. Thus we perform a classical multidimensional scaling analysis (CMS) to associate Euclidean coordinates to each PD \cite{cox08}. We use CMS to analyse geometric features not easily detectable from Fig.~\ref{fig:dm} a,b. The validity of this method is confirmed by noting that Euclidean distances between PDs are minimally distorted with respect to the corresponding bottleneck distances (see SM). We call the PD orbit under the CMS the \emph{topological pathway} (ToP).  In Fig.~\ref{fig:dm}~c we show the 3D projections of the nematic-isotropic (red) and isotropic-nematic (blue) ToPs: the red ToP is bigger than the blue ToP. This is explained as follows: the nanoparticles travel longer distances when the systems is heated than when it is cooled \cite{Lesiak2019}. Since the dynamics of the whole system depends on the dynamics of the nanoparticles, the longer the distances travelled by these, the bigger the observed distances between PDs. 

\begin{figure}
  \centering
  \includegraphics[scale=1.0]{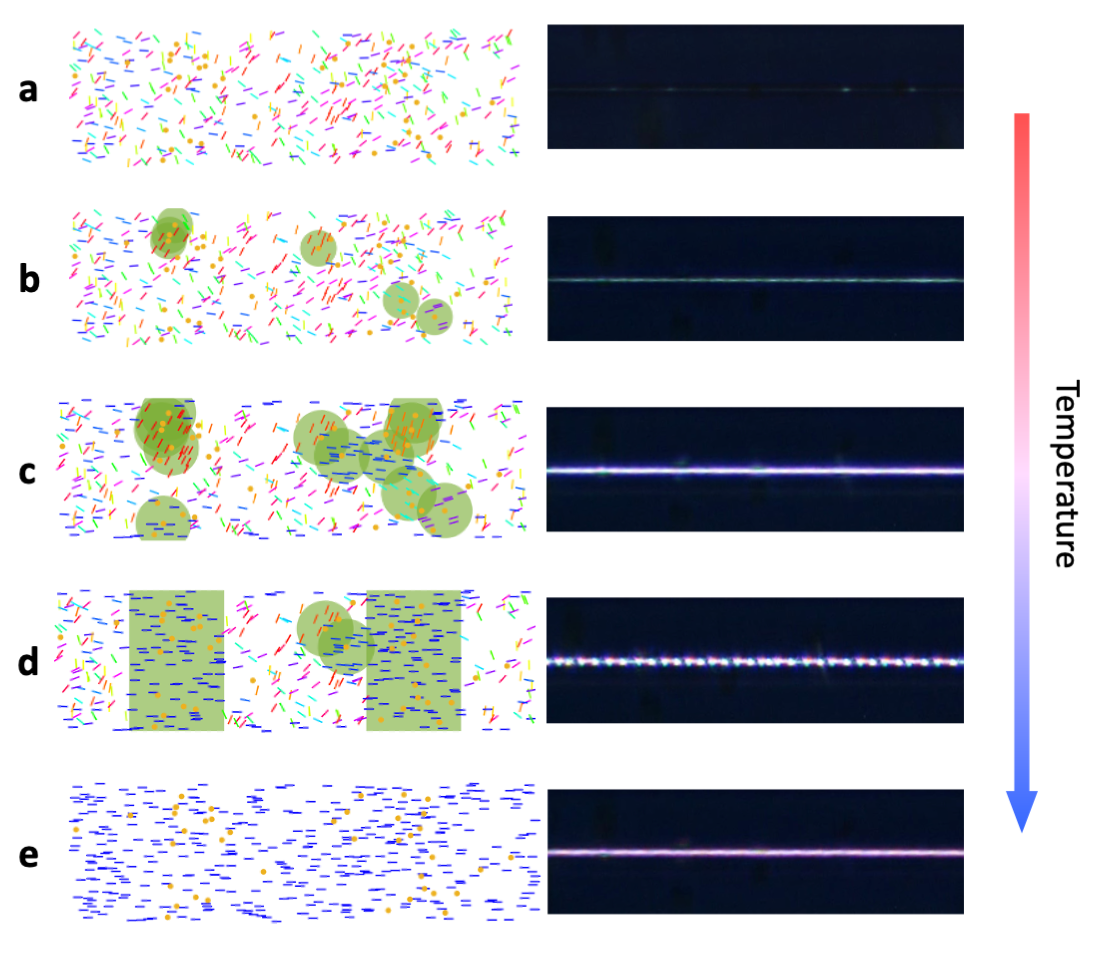}
 \caption{\textbf{Schematic representation of the isotropic-nematic phase transition and video frames of the \SI{6}{\micro m} capillary during its cooling, with the temperature decreasing from top to bottom.} \textbf{a,} Isotropic state of the LC nanocomposite; the bars represent the colour coded local orientation of the LC molecules and the yellow circles the gold nanospheres \textbf{b-d,} Formation and growth of nematic domains in the nanoparticle-poor and nanoparticle-rich regions (green areas), with a time delay. \textbf{e,} Homogeneous nematic state of the LC nanocomposite.}
 \label{fig:Iso-N}
\end{figure}

The ToP of the nematic-isotropic process shows a loop-like trajectory (Fig.~\ref{fig:dm}d), whereas the isotropic-nematic process shows two (Fig.~\ref{fig:dm}e). A loop-like shape indicates that at the beginning of the phase transition, the topology of the system starts changing, and moves away from its starting point. Once the SH reaches its maximum, the system moves back close to its starting point. Moreover we notice that the number of loops in the ToPs account for the number of thick rays coming out from the diagonal in Fig.~\ref{fig:dm} a,b and the number of local maxima in Fig.~\ref{fig:norm} a,b.
 All these results lead to the conclusion that the isotropic-nematic phase transition should be considered as a two-stage process. On cooling the sample, the phase transition to the nematic phase occurs first in the nanoparticle-rich regions (Fig.~\ref{fig:Iso-N}b). The phase transition in these regions ends at the beginning of the phase transition in the nanoparticle-poor regions (Fig.~\ref{fig:Iso-N}c). The second stage phase of the transition process is smoother (Fig.~\ref{fig:Iso-N}~d-e). It may be influenced by dispersed nanoparticles, which initiate the phase transition process (Fig.~\ref{fig:Iso-N} c). 

In the next section we present an algorithmic quantitative analysis that is particularly powerful when visual inspection is not practical.  For that purpose we measure the \emph{velocity} and \emph{curvature} of the ToP.
 
\begin{figure}
  \centering
  \includegraphics[scale=0.54]{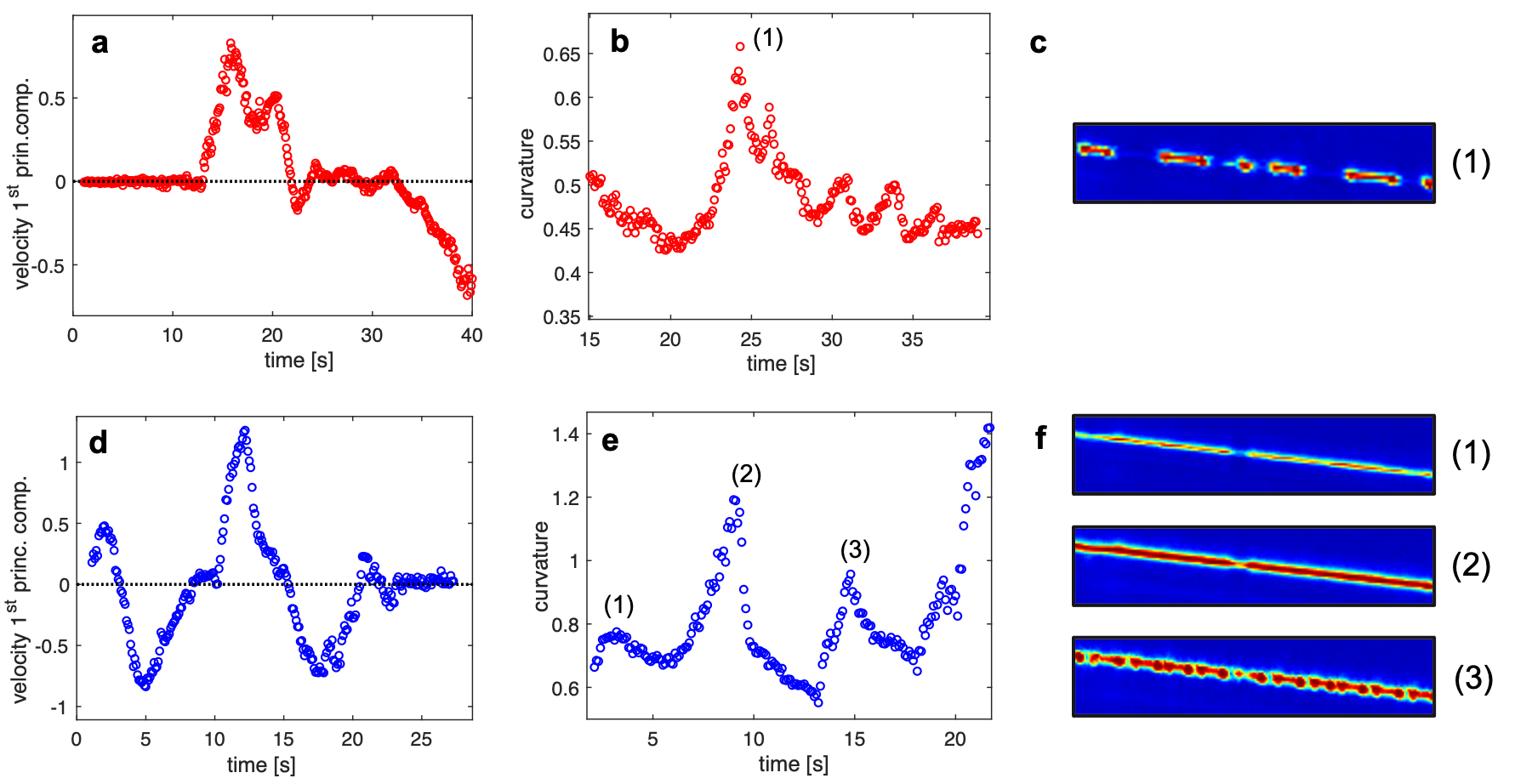}
  \caption{\textbf{The top and bottom rows represent the nematic-isotropic and isotropic-nematic transitions, respectively, for the \SI{6}{\micro m} capillary.} \textbf{a,d,} The velocity of the ToPs, while \textbf{b,e,} -  their curvature. \textbf{c,f,} False colour images of the capillary, with light intensity increasing from blue (total darkness) to red (maximum brightness).  The maximum of curvature in (\textbf{b}), indicated by (\textbf{1}), corresponds to the transition point when the SH reaches its maximum and the nematic domains begin to collapse (\textbf{c}).  The three curvature maxima in (\textbf{e}) correspond to the same-numbered pictures in (\textbf{f}): (\textbf{1}) is the early start of nematic domain formation; (\textbf{2}) their further development; (\textbf{3}) the early start of the ordered nematic phase. These three stages of the dynamics are also illustrated schematically in Fig.~\ref{fig:Iso-N}~b-d.}
  \label{fig:vel_curv}
\end{figure}

\section*{Algorithmic analysis of the system dynamics}

The \emph{pathway velocity} $v_{t}$ measures the rate of change of the system topology: 
\[
v_t = \frac{1}{2} \left ( \mathbf{x}_{(t+1)\mathrm{s}} - \mathbf x_{(t-1)\mathrm{s}} \right ),
\]
where $\mathbf{x}_{t\,\mathrm{s}}$ is the first coordinate of the ToP at time $t$~s.  This choice has been made for convenience of plotting and to preserve the directional information.

The \emph{pathway curvature} $k_t$ measures how much the ToP is bending at a given time of $t$~s.  It is defined as the inverse of the radius of the circle interpolating $\mathbf{x}_{(t-2)\mathrm{s}}$, $\mathbf{x}_{t\,\mathrm{s}}$ and $\mathbf{x}_{(t+2)\mathrm{s}}$, where $\mathbf{x}_{t\,\mathrm{s}}$ are the Euclidean coordinates of the ToP at time $t$~s. We use these two numerical descriptors to identify, in an algorithmic manner, physical states that are critical in the evolution of the soft nanocomposite.

The velocity and curvature of the nematic-isotropic and isotropic-nematic ToPs are shown in Fig.~\ref{fig:vel_curv}. This figure should be compared with the SH and distance matrix plots in Fig.~\ref{fig:norm}~a,b and Fig.~\ref{fig:dm}~a,b, respectively. We first consider the heating process. The formation of isotropic bubbles leads to a rapid increase of $v_t$, Fig.~\ref{fig:vel_curv}~a. As new isotropic bubbles form, the velocity increases, reaching a maximum; during the N-Iso phase, when only the size but not the number of the domains changes, $v_t$ decreases to zero. Towards the end of the phase transition, the number of domains decreases and $v_t$ grows in absolute value. The curvature plot (Fig.~\ref{fig:vel_curv}~b) supports this observation, having its maximum value in the state with the highest SH, Fig.~\ref{fig:vel_curv}~c.

The isotropic-nematic process shows a different behaviour, Fig.~\ref{fig:vel_curv}~d-f. There are two time periods when $v_t$ becomes negative in agreement with  the two loop-like trajectories in Fig.~\ref{fig:dm}~e. Furthermore, $v_t$ is zero at 3 different isolated points that are also critical points of the SH plot. The local maxima of the pathway curvature, Fig.~\ref{fig:vel_curv}~e, correspond to  nematic domain formation around the nanospheres in the first stage of the phase transition, Fig.~\ref{fig:vel_curv}~f(1), a transition state with a minimum in the SH, Fig.~\ref{fig:vel_curv}~f(2), and a state in the second stage of the phase transition with maximum SH, Fig.~\ref{fig:vel_curv}~f(3).  

To conclude this analysis, we point out that ToPs might probe the energy landscape on which the dynamics of the phase transitions under nonequilibrium conditions takes place: zero velocity points and associated topological configurations can be considered as metastable states of the ToPs, reachable through connecting saddle states with activation barriers. Although the nematic-isotropic phase transition is considered in a number of theories~\cite{Singh00}, to the best of our knowledge, the theories of mesoscopic nonequilibrium thermodynamics~\cite{Cugliandolo13} have not yet been applied to its study. An evolution model has been recently demonstrated to interpret complex phenomenology in disordered glasses~\cite{Fan17}. The analysis presented in this paper can potentially be used to provide the necessary experimental details for the development of a mesoscopic model of nonequilibrium phase transitions in nematic LC-based systems, which could be based on the Landau-de Gennes theory~\cite{Gramsbergen86} combined with multiscale nonequilibrium thermodynamics~\cite{Reguera05, Grmela13}. 

\section*{Outlook}

\begin{figure}
  \centering
  \includegraphics[scale=0.63]{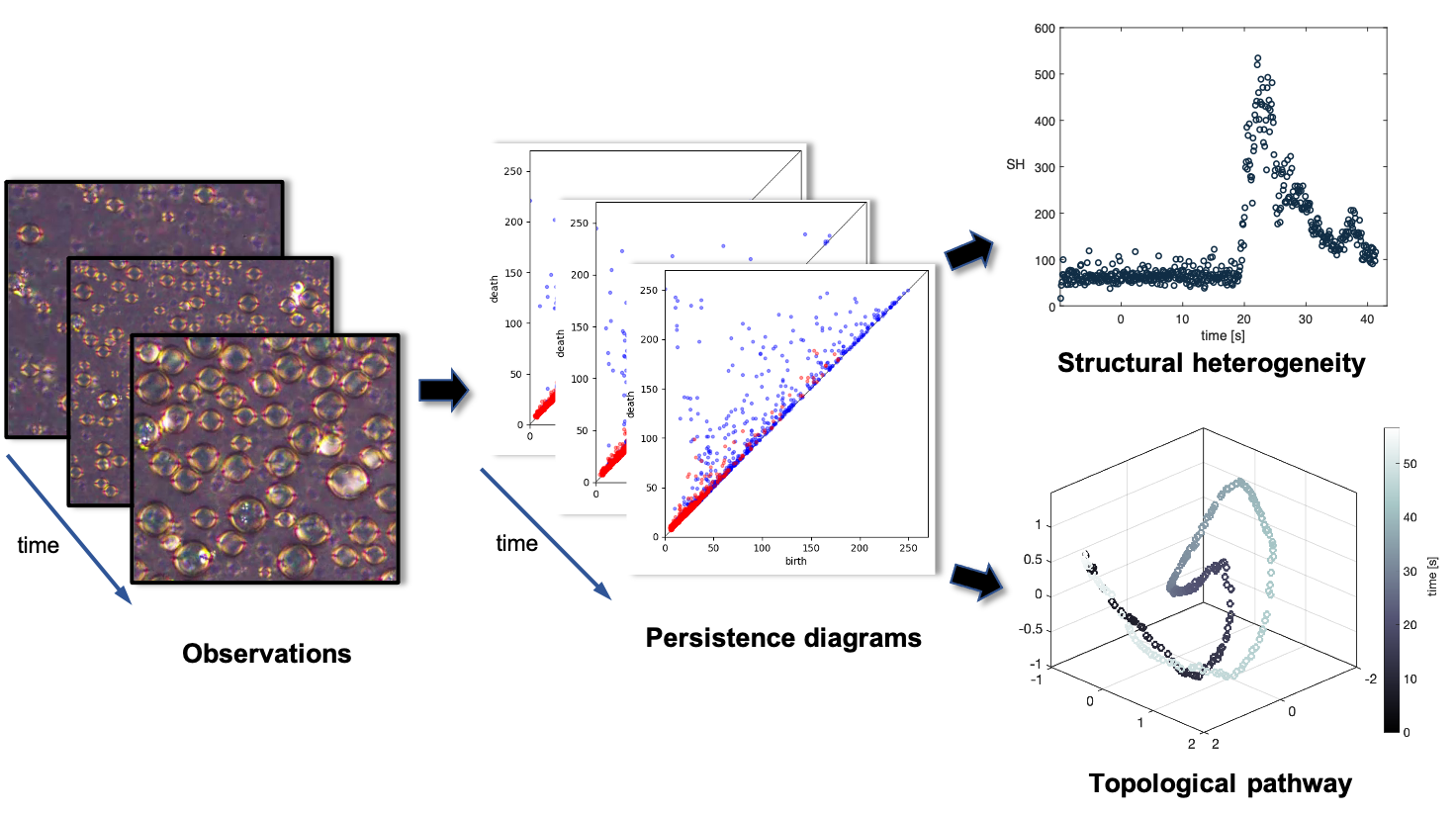}
  \caption{\textbf{Framework for the topological characterisation of soft matter systems.} The evolution of the physical system is captured using time-dependent persistent homology, which is summarised through structural heterogeneity and the ToP.  }
  \label{fig:workflow}
\end{figure}

\emph{Structural heterogeneity} (SH) is a robust topological characteristic that can  \emph{qualitatively} and \emph{quantitatively} detect subtle variations in the physical state of a soft matter system as it undergoes a dynamical process. In the particular case of the system analysed in this work, SH detects a hysteresis behaviour in its phase transitions and reveals a two stage process for the isotropic-nematic phase transition. Additional tools, such as the topological pathway, opens up the possibility of an algorithmic study of the system dynamics: the critical points of the pathway, which can be easily detected numerically either in terms of velocity or curvature, correspond to moments of significant topological and, hence, physical upheaval for the system.  

An important outcome of this study is a framework for the quantitative analysis of complex systems. As shown in Fig.~\ref{fig:workflow}, the framework starts from experimental images of the time-evolving system, followed by the acquisition of their persistence diagrams.  With this new data one can compute the SH and the topological pathway, from which we can analyse physical features of the system. This procedure can be automated and applied to large data sets for which visual analysis may not be practical, as well as to systems where the experimental information is not limited to images and may also include biological, chemical and structural information \cite{topaz2015,xia2018,pike2020}.
It is straightforward to implement our new analytical framework in the study of other 1D and 2D soft matter systems. Our methods can also be extended to 3D systems, however, this would require complex data acquisition methods and significant computational resources.

\section*{Methods}

\noindent
\textbf{Image analysis and data processing} - We analysed video data of nematic-isotropic and isotropic-nematic phase transitions for capillaries of different diameters filled with LC nanocomposites reported earlier in \cite{Lesiak2019} (see  SM 
for experimental details).  The video frames were transformed to grey scale in MATLAB \cite{matlab} using the function \textit{rgb2gray}. We used the open source GUDHI \cite{gudhi} to compute the persistent diagrams of the video frames, and all pairwise bottleneck distances in dimension 1 (for details on the computations see \cite{gudhi:cc,gudhi:bd}). The input data in the computations of the bottleneck distance matrix was the corresponding to the capillaries of 6 and 20$\mu$m for both the nematic-isotropic and the isotropic-nematic phase transition.  All persistence diagrams were normalised in the same way for each video according to a scaling factor (see SM for details on the normalisation procedure). 
The ToPs corresponding to the bottleneck distance matrices were obtained using classical multidimensional scaling (see SM 
for details) \textit{via} the Matlab function \textit{cmdscale}. The time dependent curvature values of the ToPs were computed using an implemented code in Matlab (see SM 
for details on the curvature computation).

\section*{Acknowledgements}
This work was supported by the Leverhulme Trust (grant RPG-2019-055). Experimental studies were funded by FOTECH-1 project (WUT, Excellence Initiative: Research University (ID-UB)).

\section*{Authors' contributions}
IMS performed the Topological Data Analysis. TO analysed and interpreted the experimental results with input from TW, PL and KB, who also provided the experimental details and data.  MK, JB and GD conceived the original idea and planned the core research. MK directed the study. The paper was written by IMS, TO, JB, MK and GD, with inputs from all the authors.

\section*{Competing interests}
The authors declare that they have no conflict of interest.

\section*{Data availability}
The data sets generated and analysed during the current study are available from the corresponding author on reasonable request.

\section*{Code availability}
The code used during the current study is available from the corresponding author on reasonable request.

\clearpage
\begin{center}
\LARGE
Supplementary material

\textit{Structural heterogeneity}: a topological characteristic to track the time evolution of soft matter systems

\bigskip\bigskip

\large
Ingrid~Membrillo~Solis, Tetiana~Orlova, Karolina~Bednarska, Piotr~Lesiak, Tomasz~R.~Woli\'{n}ski, Giampaolo~D'Alessandro, Jacek~Brodzki, Malgosia~Kaczmarek

\bigskip\bigskip
\end{center}

\renewcommand{\thesection}{SM\arabic{section}}
\renewcommand{\thefigure}{S\arabic{figure}}
\setcounter{section}{0}
\setcounter{figure}{0}

\section{Experimental study}
\label{sec:experiment}
\begin{figure}[h]
  \centering
  \includegraphics[width=0.8\textwidth]{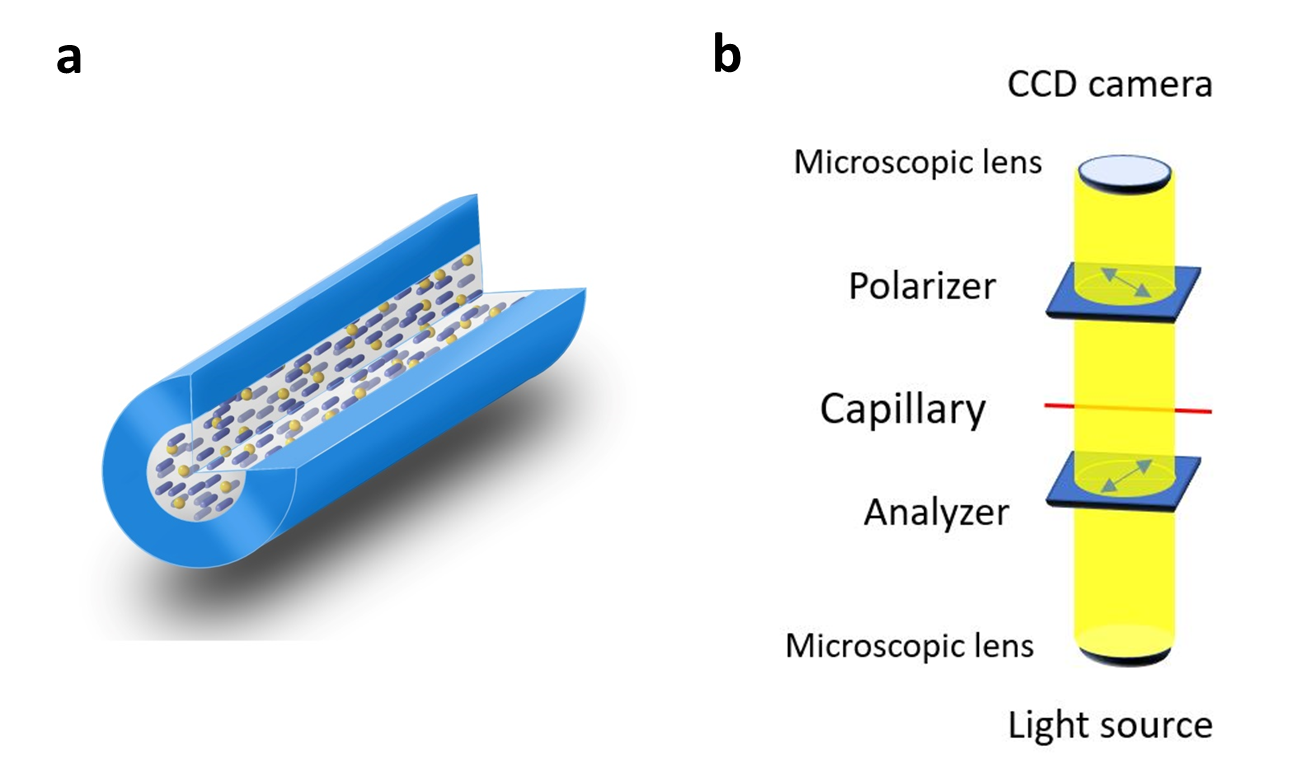}
  \caption{\textbf{Schematic diagrams of the experimental study.} \textbf {a,} LC nanocomposite filled into a cylindrical capillary. \textbf {b,} Experimental setup for collecting optical microscopy data on the structural organization of a LC nanocomposite at the mesoscopic level.}
  \label{fig:schemes}
\end{figure}

In the experimental study \cite{Lesiak2019}, LC nanocomposite was a mixture of the nematic liquid crystal 4-(\textit{trans}-4\textit{'}-\textit{n}-hexylcyclohexyl)-isothiocyanatobenzene (6CHBT) doped with gold nanoparticles ($2.5 \pm 0.4$ nm diameter for the Au core), which were covered with the promesogenic
ligand N,N-dioctyl-4-[(4\textit{'}-(10-mercaptodecyloxy)-biphenyl-4-ylo)xymethyl]benzamide (2NC8). The LC nanocomposite was filled into silica capillaries of the inner diameter 6 -- 60 $\mu$m with planar anchoring conditions at the glass-LC interface through
capillary action (Fig.~\ref{fig:schemes}~a). 

 To observe the order-disorder phase transitions, the capillaries were placed in a Nikon Eclipse polarized optical microscope between crossed polarizers at the angle $45^{\circ}$ to their axes, which ensures the appearance of nematic and isotropic domains as bright and dark, respectively (Fig.~\ref{fig:schemes}~b). The Linkam THMS600 microscope temperature stage with accuracy of \SI{0.1}{\celsius} was used for temperature variations. The capillaries were heated and cooled at the same rate of \SI{1}{\celsius/min}. All observations were recorded by using the Nikon DS-Fi2 CCD camera with video frame size 1280/960, spatial resolution \SI{0.9}{\micro m} per pixel and frame rate 50 fps. 

\section{Persistent homology of cubical complexes}
\label{sec:persistent-homology}

We present a brief introduction to the theory persistent homology for cubical complexes and give the mathematical background to the persistence diagrams in Fig.~1. For a more detailed exposition of this method and their applications we refer the reader to \cite{kac2006,Otter17,oudot2015}.

An elementary interval is a closed interval $I\subset \mathbb R$ of one of the following forms: a) $[k,k+1]$, called 1-interval or edge or b) $[k,k]$ called 0-interval or vertex, for some $k\in\mathbb Z$. An elementary cube of dimension $p$ or $p$-cube $\omega$ is a product of finitely many elementary intervals with $p$ 1-intervals as factors. The boundary of a $p$-cube $\omega$ is the union of all $p$-1-cubes $\sigma$ such that $\sigma\in\omega$. A set $X\subset \mathbb R^d$ is cubical 
if $X$ is the union of finitely many elementary cubes, $X=\bigcup_{i=1}^r\omega_{i}$. A cubical complex is a cubical set $X\subset \mathbb R^d$ topologized with the subspace topology of $\mathbb R^d$. For a field $R$, a cubical $p$-chain $Q$ of a cubical set $X$ is a finite sum $Q=\sum_i a_i\omega_i$, where each $\omega_i$ is an elementary $p$-cube contained in $X$ and $a_i\in R$. The set of all cubical $p$-chains on $X$, denoted $Q_p(X)$, has the structure of a vector space over $R$ of dimension equal to the number of elementary $p$-cubes contained in $X$. The boundary map $\partial_p:Q_p(X)\to Q_{p-1}(X)$ is a map on $p$-chains. This map is defined on an elementary $p$-cube $\omega = I_1 \times \cdots \times I_d$ as 
$\partial_p \omega=\sum_{i=1}^n(-1)^{i-1} I_1\times\cdots\times {\partial} I_i\times\cdots I_d,$ where $\partial I_i=0$, if $I_i=[k,k]$, and $\partial I_i=[k+1,k+1]-[k,k]$, if $I_i=[k+1,k]$, $k\in\mathbb Z$. 
Note that the boundary map is defined so that a $p$-chain is mapped to its boundary and satisfies $\partial_p\circ\partial_{p+1}=0$.

Let $K(X)$ be a cubical complex and $R$ be a field  The \textit{\textbf{cubical $p$-th homology}} of $X$ with coefficients in $R$ is given 
as \[H_p(X,R):=\frac{kernel (\partial_p)}{image(\partial_{p-1})}.\]
 A filtration $\mathcal F(X)$ of a cubical complex $X$ is a collection of sets $\{X_i\}_{i=1}^n$, $n\in \mathbb N$, such that every set $X_i$ is a cubical complex and $X_1\subset X_2\subset\cdots\subset X_n=X$. 
 
 Given a filtration of a cubical complex,  $\mathcal F(X)=\{X_i\}_{i=1}^n$, the \textit{\textbf{$p$-th persistent homology}} of $\mathcal F(X)$ is a pair
\[
\big( \{H_p(X_i,R)\}_{1 \leq i \leq n}, \{ f_{i,j} \}_{1 \leq i \leq j \leq n} \big)
\]
where $H_p(X_i,R)$ is the homology vector space of $X_i$ with coefficients in $R$, $f_{i,j}^p: H_p(X_i) \to H_p(X_j)$ is the map induced by the inclusion $X_i \hookrightarrow X_j$ for $1 \leq i \leq j \leq n$. The dimension of the image of $f_{i,j}^p$ is called the $p$-th persistent Betti numbers $\beta_p^{i,j}$. Betti numbers account for the number of topological features or $p$-cycles. For instance, the zeroth Betti number  $\beta_0^{i,j}$ counts $0$-cycles or connected components, the first Betti number $\beta_1^{i,j}$ counts 1-cycles or holes and the second Betti number $\beta_2^{i,j}$ counts 2-cycles or voids. The Fundamental Theorem of Persistent homology \cite{zom05} allows one to characterise the persistent homology of a filtration of a space $X$
as a \textit{persistence diagram}  which is a collection of pairs $\{(b_i,d_i )\}_{i=1}$, where $1 \leq b_i \leq d_i \leq n$.  These are, for example, the diagrams in Fig. 1 of the main text.

\section{Normalised persistence diagrams}
\label{sec:image-processing}

Given a gray scale video with frames $w$ pixels wide and $h$ pixels high, we let $\mathcal{I}_j$ be its $j$-th frame with the associated intensity function $f_j: \{1,\ldots,w\} \times \{1,\ldots,h\} \to \{0,\ldots,255\}$, and we define a filtration $\mathcal{F}(\mathcal{I}_j) = \{ X_{j,i} \}_{i=0}^{255}$, as above. We also let $D_j$ be the persistence diagram associated with $\mathcal{F}(\mathcal{I}_j)$. 
We define a \textit{scaling factor} for the video frames by the formula

\begin{equation}
s := \max_j \left( \frac{1}{wh} \sum_{\substack{m=1,w \\ n=1,h}} f_{j}(m,n) \right),
\end{equation}
that is, $s$ is the maximal average pixel intensity among all the frames in the recording. We use this definition to rescale the persistence diagrams we obtained: for each we define a \textit{normalised} persistence diagram $D_j'$ as the multiset $\{ (x/s,y/s) \mid (x,y) \in D_j \}$. 

\section{Multidimensional scaling}
\label{sec:multidimensional-scaling}
Starting from a $n$-by-$n$ matrix of Euclidean distances between $n$ points, the classical multidimensional scaling algorithm finds coordinates for the points in such a way that distances are preserved: given a data set $\mathcal S$ with $n$ points in a $k$-dimensional Euclidean space, let the coordinates $x_j$ of the $ith$-point given by $x_j=(x_{k1},\dots,x_{kn})^T$. The Euclidean distance between two points $x_p,x_q\in\mathcal S$ is given by  $d^2(x_p,x_q)=(x_p-x_q)^T(x_p-x_q)$. Let $A$ be the inner product matrix, where
$(A)_{pq}=x_p^T\cdot x_q$. 
The matrix $A$ is found using the squared distances $\{d({x_p,x_q})\}$ and from $A$ one can find the coordinates of the points in $\mathcal S$ up to a choice of orientation for each coordinate.   
The inner product matrix $A$ is positive semidefinite. In this case classical multidimensional scaling coincides with principal component analysis. Therefore, the eigenvalues of the inner product matrix are positive. In many applications, however, the initial matrix of distances might not be constructed using Euclidean distance, or even more, it might only be a matrix of dissimilarities, nor of distances. In those cases, the algorithm still can find Euclidean coordinates for the points, although the eigenvalues of the matrix $A$ might be negative. If the absolute values of the negative eigenvalues are relatively small compared to the positive eigenvalues, multidimensional scaling still can give a fair linear approximation of the dissimilarity or distance matrix using the principal coordinates associated with the largest positive eigenvalues. A detailed exposition of the algorithm to obtain the Euclidean coordinates using classical multidimensional scaling can be found in \cite{cox08}.

\begin{figure}
  \centering
  \includegraphics[width=1\textwidth]{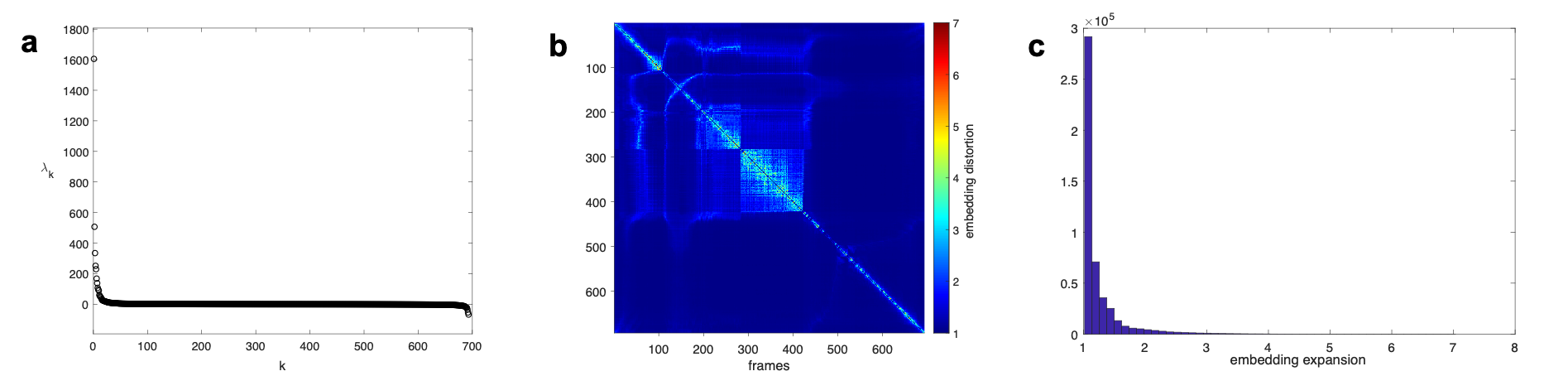}
  \caption{\textbf{Eigenvalues and embedding expansion of the classical multidimensional scaling method.} \textbf {a,} Eigenvalues $\lambda_k$ of the inner product matrix. \textbf{b,} Embedding expansion matrix; the largest metric distortion occurs near the diagonal, that is, in points very close to each another. \textbf {c,} The histogram of the embedding expansion shows that most of the distance between persistence diagrams are well approximated by the classical multidimensional scaling. 
  }
  \label{fig:distorsion6h}
\end{figure}

The embedding expansion $\mathcal E$ of the classical mutidimensional scaling $\rho$ is defined as 
$$\mathcal E(\rho):=\max_{x,y\in X}\frac{d(\rho(x),\rho(y)}{bd(x,y))} $$
where $d$ is the Euclidean distance and $bd$ is the bottleneck distance. The value of the embedding expansion allows one to measure how much the distances between points are distorted in the Euclidean embedding with respect to the bottleneck distances. If the classical multidimensional scaling $\rho$ is an isometry, i.e. pairwise distances are preserved, then $\mathcal E(\rho)=1$. 
It is known that the distortion of the bottleneck distance matrix by any embedding will depend on the number of points in the persistence diagrams \cite{carriere2019}.  In order to determine the suitability of the classical multidimensional scaling for the analysis of the videos frames of the N-I and I-N phase transition in a \SI{6}{\micro m}-capillary, we analyse the eigenvalues of the inner product matrix and the embedding expansion $\mathcal E$ obtained from the classical multidimensional scaling algorithm.
The negative eigenvalues  of the inner product matrix are small in absolute value compared to the biggest positive ones (Fig. \ref{fig:distorsion6h} (a)). The values of $\mathcal E$ are small for most pairs of persistence diagrams as it is shown in Fig. \ref{fig:distorsion6h} (b-c).  

\section{Curvature in topological pathways}
\label{sec:curvature}

A topological pathway is realised as a curve $\gamma:[0,T] \to \mathbb{R}^n$ parametrised by time. We fix a constant $\delta > 0$, and approximate the radius of curvature, $r_t$, of the topological pathway at time $t \in [\delta,T-\delta]$ as the radius of the circle passing through the points $\gamma(t-\delta)$, $\gamma(t)$ and $\gamma(t+\delta)$. In particular, we set
\[
r_t = \frac{abc}{\sqrt{(a+b+c)(a+b-c)(a+c-b)(b+c-a)}},
\]
where $a = d(\gamma(t-\delta),\gamma(t))$, $b = d(\gamma(t+\delta),\gamma(t-\delta))$ and $c = d(\gamma(t+\delta),\gamma(t))$; here $d(x,y)$ is the Euclidean distance between points $x,y \in \mathbb{R}^n$ \cite{coxeter1961}. We then set the curvature at time $t$ to be
\[
k_t = \frac{1}{r_t}.
\]




\section{Additional data}
\label{sec:additional-data}

\begin{figure}
  \centering
  \includegraphics[width=\textwidth]{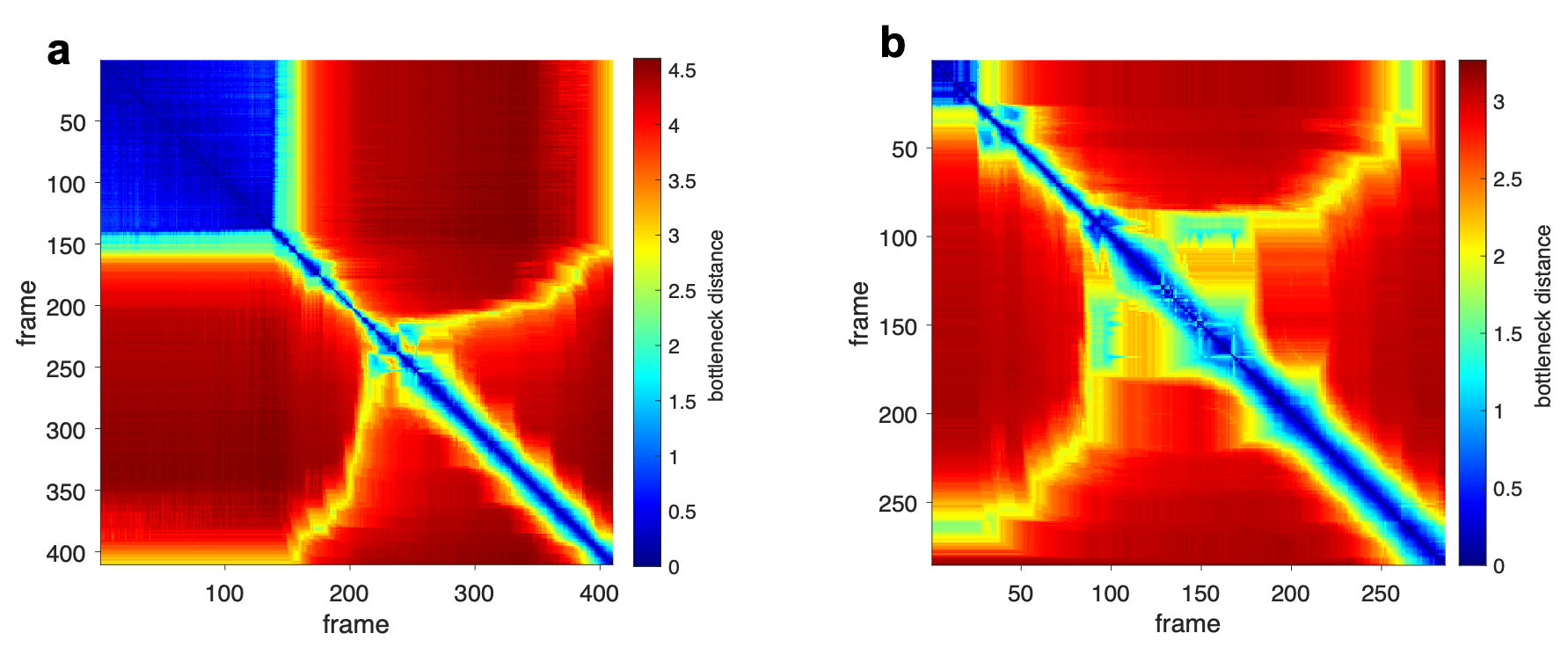}
  \caption{\textbf{Coexistence of mesophases and stability.} The distance matrices for the capillaries with diameter \textbf{a}) \SI{6}{\micro m} and \textbf{b}) \SI{20}{\micro m}. The middle yellow square corresponds to the coexistence of nematic and isotropic phases. The time of coexistence increases with the diameter of the capillary. The frame rate is 10~fps in both images.}
  \label{fig:dms}
\end{figure}

\label{sec:addit-exper-data}

\subsection{Supplementary movie 1}
Persistence diagrams and corresponding grayscale images obtained from full-colour video frames of the nematic to isotropic phase transition in the capillary of 6 $\mu$m diameter. Frame rate is 10 fps.

\subsection{Supplementary movie 1}
Persistence diagrams and corresponding grayscale images obtained from full-colour video frames of the isotropic to nematic phase transition in the capillary of 6 $\mu$m diameter. Frame rate is 10 fps. 
\subsection{Supplementary movie 3}
Persistence diagrams and corresponding grayscale images obtained from full-colour video frames of the nematic to isotropic phase transition in the capillary of 20 $\mu$m diameter. Frame rate is 10 fps.

\subsection{Coexistence of the nematic and isotropic  mesophases}
In Fig.~\ref{fig:dms} we compare the distance matrix for the nematic to isotropic phase transition in two capillaries of different diameter, namely \SI{6}{\micro m} and \SI{20}{\micro m}. The dynamics of the phase transition seem to be similar in both capillaries. The deep blue squares in the top left corner of both distance matrices correspond to point before the phase transition starts. The yellow square in the middle of both distance matrices corresponds to configurations where there is a coexistence of both nematic and isotropic phase transition. The  coexistence time is longer in the capillary with larger diameter.

\bibliographystyle{naturemag}
\bibliography{bibliography}

\end{document}